# Octupole Contributions to the Generalized Oscillator Strengths of Discrete Dipole Transitions in Noble Gases.

M. Ya. Amusia,[1,2] L. V. Chernysheva,[2] Z. Felfli,[3] and A. Z. Msezane[3]

[1]*Racah Institute of Physics, The Hebrew University, Jerusalem 91904, Israel*
[2]*A. F. Ioffe Physical- Technical Institute, St. Petersburg 194021, Russia*
[3]*Center for Theoretical Studies of Physical Systems, Clark Atlanta University, Atlanta, GA 30314*

(Dated: January 2, 2007)

The generalized oscillator strengths (GOS's) of discrete excitations $np \to nd$, both dipole ($L = 1$) and octupole ($L = 3$) are studied, the latter for the first time. We demonstrate that although the relevant transitions in the same atom are closely located in energy, the dependence of their GOS's on the momentum transfer squared $q^2$, is remarkably different, *viz.* the GOS's corresponding to $L = 3$ have at least one extra maximum as a function of $q^2$ and dominate over those of the $L = 1$, starting from about $q^2 = 1.5$ *a.u.* The calculations were performed in the one particle Hartree-Fock approximation and with account of many-electron correlations via the Random Phase Approximation with Exchange. The GOS's are studied for values of $q^2$ up to 30 *a.u.*

PACS numbers: 32.80.Gc,34.50.-s,52.20.Hv,03.65.Nk

## I. INTRODUCTION

Here we consider the lowest energy optically allowed transitions for the outer $np$ subshells of the noble gas atoms $np \to nd$ and $np \to (n+1)s$. The former series of transitions in Ne, Ar, Kr and Xe, namely $np \to nd$ can be dipole and octupole, while the latter can only be a pure dipole. The essential feature of the levels considered is that those with the same configurations $np \to nd$ but different total angular momentum $L$, $L = 1$ and $L = 3$, are closely located and hardly separable in existing experiments. This means that they will be excited by electron (or other charged particle) impact simultaneously, but decay via photon emission separately; the decay of the octupole excitation being about eight orders of magnitude slower than the dipole. As a result, soon after excitation of a gas volume by incoming charged particles, only octupole levels will survive. Therefore, separate calculations of octupole GOS's are of importance and interest, presenting the probability of populating and studying the long living excited levels.

The generalized oscillator strength, introduced by Bethe [1] and reviewed by Inokuti [2], characterizes fast electron inelastic scattering. It manifests directly the atomic wave functions and the dynamics of scattering. Since then, the GOS has received attention from a variety of perspectives: determination of the correct spectral assignment [3], exploring the excitation dynamics [4], probing the intricate nature of the valence-shell and inner-shell electron excitation [5], investigation of the influence of the angular resolution and pressure effects on the position and amplitude of the GOS minima [6], investigation of the GOS ratio [7] and various correlation effects [8-11], as well as multiple minima [12].

One of the most important utilities of the GOS concept in the limit $q^2 \to 0$ is in the determination of optical oscillator strengths (OOS's) from absolute differential cross sections (DCS's) [13-16]. Implicit in this is the extrapolation of the measured data through sometimes the unphysical region [17]. The limiting behavior of the GOS as $q^2 \to 0$ has been a subject of continuing interest because of the difficulty of measuring reliably the electron DCS's for atoms, ions and molecules at and near zero scattering angles [3,14,18]. This difficulty still plagues measurements of the DCS's [19, 20], including the most recent measurements of the DCS's [21, 22]. Thus a thorough understanding of the behavior of the GOS's near $q^2 = 0$ is imperative to guide measurements.

To place the current investigation of the GOS's for the $np - nd$ and $np - ns$ transitions in Ne, Ar, Kr and Xe in perspective, it is important to highlight some significant new manifestations that have been uncovered in the recent studies of correlation effects in the GOS's for atomic transitions:

1) Recently, GOS's for monopole, dipole and quadrupole transitions of the noble gas atoms have been investigated in both Hartree-Fock (HF) and Random Phase Approximation with Exchange (RPAE) approximations as functions of $q$ and the energy transferred, $\omega$ [8]. There it was found that electron correlations, both intra-shell and inter-shell are important in the GOS's for all values of $q$ and $\omega$ investigated and that the variation of the GOS's with $q$ and $\omega$ is characterized by maxima and minima, arising entirely from many-electron correlations. These results have been used to understand and interpret [23, 24] the first experimental observation of the absolute GOS for the nondipole Ar $3p - 4p$ transition [5]. Of great significance is that the calculated GOS's for discrete transitions permitted the determination of their multipolarity quite reliably [10]. The interpretation is of particular interest for nondipole transitions since they cannot be observed in photon absorption. Of even greater importance and accomplishment in the recent measurement of the GOS's for the valence-shell excitation of Ar is the separate measurement of the electric monopole and quadrupole of the GOS's for the valence-shell excitation of Ar [25].

2) For the outer and intermediate shells of Xe, Cs and Ba, correlations have been discovered to produce impressive manifestations of intra-doublet interchannel interaction [26], yielding new structures in nondipole parameters and GOS's [9]. GOS results for Xe, Cs and Ba demonstrated the leveraging role of the spin-orbit interaction, *viz.* the strong interaction between components of the spin-orbit doublet of the $3d$ electrons in Cs, Ba and Xe. This leads to the appearance of an additional maximum in the GOS for the $3d_{5/2}$ sub-shell, due to the action of the $3d_{3/2}$ electrons. The inter-doublet correlations were found to be very important in the monopole, dipole and quadrupole transitions.

3) Generalized oscillator strengths for monopole, dipole and quadrupole transitions in the negative ions $I^-$ and $Si^-$ have been investigated to assess the extent of importance of correlations [27]. It was found that the GOS's for monopole and dipole transitions, are generally characterized by two distinct sets of maxima as functions of $\omega$, being most pronounced for the dipole transitions. For the negative ions, there are two interesting and significant peculiarities. Firstly, contrary to the well-known behavior of atomic transitions, the limit of the GOS approaches zero as $q^2 \to 0$ for the dipole transition. Secondly, in both the monopole and the quadrupole transitions, the GOS corresponding to $q = 2$ *a.u.* starts being zero at threshold and becomes dominant beyond about $\omega = 6$ Ry.

## II. THEORY

In this paper, we consider a relatively simple case where the transition energy can be specified almost entirely by the one-electron nomenclature, namely by the principal quantum number and angular momentum of the exciting electron in its initial and final states, $nl$ and $n'l'$, respectively and by the total angular momentum of the excitation $L$. Both the energies $\omega_{nl,n'l'}$ and the GOS's are affected by the multi-electron correlations, since the one-electron approximation is very often not accurate enough, even for qualitative, not to mention quantitative, description. With the above in mind, we have performed calculations both in the one particle Hartree-Fock approximation and with account of many-electron correlations via the RPAE [28].

The RPAE has proved to be very effective in describing the photoionization cross sections and GOS's [8,23,24,28-30] including rather delicate features of the dipole and nondipole angular anisotropy parameters of photoelectrons, where impressive manifestations of electron correlations were recently observed in good accord with both experiment and calculations (see *e.g.* [26, 31-36]).

The theoretical consideration in this paper of dipole and octupole excitations is similar to that of quadrupole and monopole excitations in [23,24]. However for the convenience of the reader and for better understanding of the results we repeat here the main points of consideration of GOS in general and in the one-electron HF frame and with account of correlations in the RPAE frame. All necessary formulas are also presented.

One specific feature of the considered in this pasper discrete excitations require special attentin. Namely, in Kr there are two closely located dipole excitations: 4p-4d energy is 0.93330 Ry, while 4p-6s energy is 0.93403 Ry and 4p-5d energy is 0.98429 Ry, while 4p-7s energy is 0.98479 Ry. As to the method implemented in the computing program in [38] and used in [23,24], it is suitable for the case of a single relatively isolated discrete excitation, for which all other act as perturbation. In our case the closely located levels are distorbing each other very strong. Therefore the corresponding interaction has to be taken into account accurately enough.

The inelastic scattering cross sections of fast electrons or other charged particles incident upon atoms or molecules are expressed via the GOS $G(\omega, q)$ [1,37] which is a function of the energy $\omega$ and the momentum transferred $q$ to the target in the collision process. The GOS is defined as [1] (atomic units are used throughout this paper)

$$G_{fi}(\omega, q) = \frac{2\omega}{q^2} \mid \Sigma_{j=1}^{N} \int \psi_f^*(\vec{r_1}, ..., \vec{r_N}) exp(i\vec{q} \cdot \vec{r_j}) \psi_i(\vec{r_1}, ..., \vec{r_N}) d\vec{r_1}...d\vec{r_N} \mid^2 \quad (1)$$

where N is the number of atomic electrons and $\psi_{i,f}$ are the atomic wave functions in the initial and final states with energies $E_i$ and $E_f$, respectively, and $\omega = E_f - E_i$. Because the projectile is assumed to be fast, its wave functions are plane waves and its mass M enters the GOS indirectly, namely via the energy and momentum conservation law

$$\frac{p^2}{2M} - \frac{(\vec{p} - \vec{q})^2}{2M} = \omega \quad (2)$$

Here $\vec{p}$ is the momentum of the projectile. It follows from the GOS definition Eq. (1) that when $q = 0$ the GOS coincides with the OOS or is simply proportional to the photoionization cross section (see for example [37]), depending upon whether the final state is a discrete excitation or belongs to the continuous spectrum. The energy $\omega$ enters the GOS either via a factor in Eq. (1) or indirectly, via the energy $E_f$ of the final state $\mid f >$.

In the one-electron Hartree-Fock approximation Eq. (1) simplifies considerably, reducing to

$$g_{fi}^L(q,\omega_{fi}) = \frac{2\omega_{fi}}{q^2}|\int \phi_f^*(\vec{r})j_L(qr)P_L(cos\vartheta)\phi_i(\vec{r})d\vec{r}|^2 = \frac{2\omega_{fi}}{q^2}|<f|j_L(qr)|i>|^2 \quad (3)$$

where $\phi_{f,i}(\vec{r}) = R_{n_{f,i}}Y_{l_{f,i},m_{f,i}}(\theta_{\vec{r}},\varphi_{\vec{r}})\chi_{s_{f,i}}$ are the HF one-electron wave functions with their radial, angular and spin parts, respectively, $j_L(qr)$ is the spherical Bessel function, $P_L(cos\vartheta)$ is the Legendre polynomial and $cos\vartheta = \vec{q}\cdot\vec{r}/qr$. The excitation energy of the $i \to f$ transition is denoted as $\omega_{fi}$. The principal quantum number, the angular momentum, its projection and spin quantum numbers of the initial $i$ and final $f$ states are denoted by $n_{f,i}$, $l_{f,i}$, $m_{f,i}$ and $s_{f,i}$, respectively.

To take into account of many-electron correlations in the RPAE the following system of equations was solved

$$<f|A_L(q,\omega_{fi}^R)|i> = <f|j_L(qr)|i> + (\Sigma\int_{n'\leq F,k'>F} - \Sigma\int_{n'>F,k'\leq F})$$
$$\times \frac{<k'|A_L(q,\omega_{fi}^R)|n'><n'f|U_L|k'i>}{\omega_{fi}^R - \epsilon_{k'} + \epsilon_{n'} + i\eta(1-2n_{k'})} \quad (4)$$

Here $\leq F(>F)$ denotes occupied (vacant) HF states, $\epsilon_n$ are the one-electron HF energies, $\eta \to +0$ and $n_k = l(0)$ for $k \leq F(>F)$; $<nf|U|ki>_L = <nf|V|ki>_L - <nf|V|ik>_L$ is the $L$ component of the matrix elements of the Coulomb inter-electron interaction $V$. It is seen that the system of equations for each total angular momentum of an excitation $L$ is separate. The procedure of solving this equation is described in details in [28, 38]. Note that the excitation energy of the $i \to f$ transition in RPAE $\omega_{fi}^R$ is different from the HF value $\omega_{fi} = \epsilon_f - \epsilon_i$. The procedure of calculating $\omega_{fi}^R$ is also described in [28, 38].

A relation similar to Eq. (3) determines the GOS's in RPAE $G_{fi}^L(q,\omega_{fi}^R)$

$$G_{fi}^L(q,\omega_{fi}^R) = \frac{2\omega_{fi}^R}{q^2}|<f|A_L(q,\omega_{fi}^R)|i>|^2 \quad (5)$$

Here $<f|$ and $|i>$ are the final and initial HF states, respectively. Using these formulas the GOS's were calculated for dipole $L=1$ and octupole $L=3$ components.

The operator of the interaction between fast charged particles and atomic electrons can also be represented in another form than $\hat{A}(q) = \hat{A}^r(q) \equiv exp(i\vec{q}\cdot\vec{r})$. This is anologous to the case of photoionization and can be called *length* form. The other one is similar to the velocity form in photoionization and looks like [37]

$$\hat{A}^v(\omega,q) = [exp(i\mathbf{q}\cdot\mathbf{r})(\mathbf{q}\cdot\boldsymbol{\nabla} - \mathbf{q}\cdot\overleftarrow{\boldsymbol{\nabla}})exp(i\mathbf{q}\cdot\mathbf{r})] \quad (6)$$

where the upper arrow in $\overleftarrow{\boldsymbol{\nabla}}$ in Eq. (6) implies that the function standing to the left is being operated on.

For the specific case considered in this paper the explicit HF energies are $\omega_{np \to nd,(n+1)s} \equiv \epsilon_{nd,(n+1)s} - \epsilon_{np}$ with $\epsilon_{np}$, $\epsilon_{nd}$ and $\epsilon_{(n+1)s}$ being the HF one-electron energies. $R_{np}(r)$, $R_{nd}(r)$ and $R_{(n+1)s}(r)$ are the radial parts of the one-electron wave functions in the HF approximation and $L$ is the total angular momentum of the excitation, where in our case $L=1$ or 3, $n=2;3;4$ and 5 for Ne, Ar, Kr and Xe, respectively. Symbolically, the RPAE equations can be presented as [28,38]

$$\hat{T} = \hat{t} + \hat{T}\hat{\chi}U \quad (7)$$

$$t_{np \to nd,(n+1)s}^L(q) \equiv <np|\hat{t}|nd,(n+1)s> = \int R_{np}(r)j_L(qr)R_{nd,(n+1)s}(r)dr, \quad (8)$$

$U$ is the Coulomb interelectron interaction, and

$$\hat{\chi} = \hat{1}/(\omega - \omega' + i\gamma) - \hat{1}/(\omega + \omega') \quad (9)$$

with $\gamma \to 0$ and $\omega$ being the excitation energy parameter of the relevant discrete excitation, while $\omega'$ is the energy of any other, including the considered discrete or continuous spectrum excitation of another electron, which is excited by the incoming electron. Its interaction via the potential $U$ leads to the excitation of a given atom under consideration.

The RPAE values for the GOS's $F_{np \to nd,(n+1)s}(q)$ are connected to the matrix elements of $\hat{T}$ similar to the connection of $f_{np \to nd,(n+1)s}$, the HF GOS values, with $\hat{t}$. However, a complication arises for discrete excitations. This results from the fact that one of the intermediate discrete excitation energies consistent with the energy of the excitation of the level under investigation and the corresponding element of $\hat{\chi}$ becomes infinite. To circumvent this singularity an effective interaction matrix has to be created [28,38]:

$$\widehat{\widetilde{\Gamma}} = U + U\hat{\chi}'\widehat{\widetilde{\Gamma}} \tag{10}$$

where $\hat{\chi}'$ excludes only a single term, with one of the transitions $\omega' = \omega_{np \to nd,(n+1)s}$, from summation over all intermediate states [see (4)]. Then the total matrix of the effective interaction $\hat{\Gamma}$ is determined by a simple expression:

$$\hat{\Gamma} = \widehat{\widetilde{\Gamma}}(\omega - \omega_{np \to nd,(n+1)s} - \widehat{\widetilde{\Gamma}})^{-1}. \tag{11}$$

This is correct only if the interaction between two adjaisent levels is weak enough and can be accounted for perturbatively. Then instead of Eq. (3), one can arrive at the following expression for $\hat{T}$

$$\hat{T} = \hat{t} + \hat{t}\hat{\chi}\hat{\Gamma} \tag{12}$$

With the help of Eq. (11) we derive the GOS value in RPAE as

$$F_{np \to nd,(n+1)s}(q) = Z_{np \to nd,(n+1)s}\frac{2\pi\,\omega_{np \to nd,(n+1)s}}{q^2}\,|<np\,|\,\hat{T}\,|\,nd,(n+1)s>|^2 \tag{13}$$

$$\omega_{np \to nd,(n+1)s} = \epsilon_{nd,(n+1)s} - \epsilon_{np} + \widetilde{\Gamma}_{np \to nd,(n+1)s} \tag{14}$$

$$Z_{np \to nd,(n+1)s} = \left[1 - \frac{\partial\widetilde{\Gamma}_{np \to nd,(n+1)s'}}{\partial\omega}\,|_{\omega=\omega_{np \to nd,(n+1)s}}\right]^{-1}. \tag{15}$$

Here $Z_{np \to nd,(n+1)s}$ is the spectroscopic factor of the discrete excitation level.

The equations (10)-(14) determine the RPAE values for both the GOS's and the discrete excitation energies, while Eq.(3) gives the HF GOS, represented simply as $f_{np \to nd,(n+1)s}$ with the appropriate $\omega_{fi}$ used.

However, as it was mentioned above, at least in Kr one has two very close levels. In this case one had at first to introduce an auhilary matrix of effective interelectron interaction $\hat{\Gamma}_{\alpha\beta}$ that is a solution of equation similar to (10)

$$\widehat{\widetilde{\Gamma}} = U + U\hat{\chi}''\widehat{\widetilde{\Gamma}} \tag{16}$$

with $\hat{\chi}''$ that excludes two so-called "time-forward" terms, i.e. those with energy factors $\hat{1}/(\omega - \omega'' + i\gamma)$, where $\omega''$ are the energies of two strongly interacting transtions. In our case these are terms with $\omega'' = \omega_{4p \to 4d}$ and $\omega'' = \omega_{4p \to 6s}$ or $\omega'' = \omega_{4p \to 5d}$ and $\omega'' = \omega_{4p \to 7s}$. However, the relation that determines $\hat{\Gamma}$, is not that simple as (11): it is instead of being an simple algebraic became a $2 \times 2$ matrix equation.

Let us consentrate on two first levels, $4p \to 4d$ and $4p \to 6s$, denoting them as 1 and 2, respectively. In this case the equation (16) looks as

$$\begin{pmatrix} \Gamma_{11} & \Gamma_{12} \\ \Gamma_{21} & \Gamma_{22} \end{pmatrix} = \begin{pmatrix} \widetilde{\Gamma}_{11} & \widetilde{\Gamma}_{12} \\ \widetilde{\Gamma}_{21} & \widetilde{\Gamma}_{22} \end{pmatrix} + \begin{pmatrix} \widetilde{\Gamma}_{11} & \widetilde{\Gamma}_{12} \\ \widetilde{\Gamma}_{21} & \widetilde{\Gamma}_{22} \end{pmatrix} \times \begin{pmatrix} (\omega - \omega_1)^{-1} & 0 \\ 0 & (\omega - \omega_2)^{-1} \end{pmatrix} \begin{pmatrix} \Gamma_{11} & \Gamma_{12} \\ \Gamma_{21} & \Gamma_{22} \end{pmatrix} \tag{17}$$

In fact, (17) describes a two-level system that has two solutions, $\omega = \omega'_{1,2} \neq \omega_{1,2}$. Such a system was considered in application to molecules in [36]. The corresponding solution is also known:

$$\omega'_{1,2} = \frac{1}{2}(\omega_1 + \omega_2 + \widetilde{\Gamma}_{11} + \widetilde{\Gamma}_{22}) \pm \sqrt{\frac{1}{4}(\omega_1 - \omega_2 + \widetilde{\Gamma}_{11} - \widetilde{\Gamma}_{22})^2 + |\widetilde{\Gamma}_{12}|^2} \tag{18}$$

As it should be, in absence of level mixing interaction ($\widetilde{\Gamma}_{12} = 0$), $\omega'_{1,2} = \omega_{1,2}$. In principal, each level, 1 and 2 has its own spectroscopic factor $Z_{1,2}$. But since these levels are close to each other, the corresponding $Z$ values are close to each other and can be determined by (15) with $\widehat{\widetilde{\Gamma}}$ taken from (16).

## III. RESULTS OF CALCULATIONS

The calculations were performed numerically using the programs and procedures described in [38] and for the case of Kr corrected with accord of the formulas (16-18). The results of the calculations are presented in the Table 1 and the figures 1-8 below. It is important to bear in mind that at small $q^2 (q^2 \to 0)$ the dipole GOS is absolutely dominant, since the GOS dipole component is non zero at $q^2 \to 0$, corresponding to the OOS. However, with increasing $q^2$ the situation very fast changes considerably and in some cases even dramatically, since $f^{L=1}_{np \to nd,(n+1)s}(q)$ rapidly decreases with increasing $q^2$ while $f^{L=3}_{np \to nd,(n+1)s}(q)$ rapidly increases as $q^2$ at least for small $q$. Then $f^{L=3}_{np \to nd,(n+1)s}(q)$ has to reach its maximum with subsequent decrease. The $q^2$- dependence of $f$ proved to be more complicated, exhibiting maxima for all the cases considered.

Note that the results for the dipole components of the $3p \to 3d$ level in Ar were obtained earlier [23]. Here our previous results are complemented by those from the octupole contributions, calculated for the first time, to our knowledge.

## IV. ACKNOWLEDGEMENTS

MYaA is grateful for support of this research by the Israeli Science Foundation under the grant 174/03 and by the Hebrew University Intramural fund.

---

[1] H.A. Bethe, Ann. Phys. **5**, 325 (1930).
[2] M. Inokuti, Rev. Mod. Phys. **43**, 297 (1971).
[3] C. C. Turci, J. T. Francis, T. Tyliszczak, G. G. B. de Souza and A. P. Hitchcock, Phys. Rev. A **52**, 4678 (1995).
[4] X. J. Liu, L. F. Zhu, Z. S. Yuan, W. B. Li, H. D. Cheng, Y. P. Huang, Z. P. Zhong, K. Z. Xu and J. M. Li, Phys. Rev. Lett. **91** 193203 (2003),
[5] X. W. Fan and K. T. Leung, Phys. Rev. A **62**, 062703 (2000).
[6] W. B. Li, L. F. Zhu, X. J. Liu, Z. S. Yuan, J. M. Sun, H. D. Cheng, Z. P. Zhong, and K. Z. Xu, Phys. Rev. A **67**, 062708 (2003)
[7] H. D. Cheng, L. F. Zhu, Z . S. Yuan, X. J. Liu, J. M. Sun, W. C. Jiang and K. Z. Xu, Phys. Rev. A **72**, 012715 (2005)
[8] M. Ya. Amusia, A. S. Baltenkov, L. V. Chernysheva, Z. Felfli, and A. Z. Msezane, Phys. Rev. A **64**, 032711 (2001) and references therein
[9] M. Ya. Amusia, L. V. Chernysheva, Z. Felfli, and A. Z. Msezane, Phys. Rev. A **73**, 062716 (2006) and references therein
[10] M. Ya. Amusia, L. V. Chernysheva, Z. Felfli, and A. Z. Msezane, Phys. Rev. A **67**, 022703 (2003)
[11] Zhifan Chen and A.Z. Msezane, Phys. Rev. A **70**, 032714 (2004) and references therein
[12] N. Avdonina, D. Fursa, A.Z. Msezane and R. H. Pratt, Phys. Rev. A**71**, 062711 (2005)
[13] T. Y. Suzuki, Y. Sakai, B. S. Min, T. Takayanagi, K. Wakiya, H. Suzuki, T. Inaba and H. Takuma, Phys. Rev. A **43**, 5867 (1991),
[14] T. Ester and K. Kessler, J. Phys B **27** 4295 (1994)
[15] T. Y. Suzuki, H. Suzuki, S. Ohtani, B. S. Min, T. Takayanagi and K. Wakiya, Phys. Rev. A **49**, 4578 (1994)
[16] K. Z. Xu, R. F. Feng, S. L. Wu, Q. Ji, X. J. Zhang, Z. P. Feng Zhong and Y. Zheng, Phys. Rev. A. **53** 3081 (1996)
[17] Z. Felfli, A. Z. Msezane and D. Bessis, Phys. Rev. Lett. **81** 963 (1998),
[18] I. D. Williams, A. Chutjian and R. J. Mawhorter, J. Phys B **19** 2189 (1986)
[19] V. Karaganov, I. Bray and P. J. O Teubner, Phys. Rev. A **59**, 4407 (1999)
[20] M. A. Khakoo, M. Larsen, B. Paolini, X. Guo, I. Bray, A. Stelbovics, I. Kanik, S. Trajmar and G. K. James Phys. Rev. A **61**, 012701 (2000)
[21] B. Predojevic, D. Sevic, V. Pejcev, B. P. Marinkovic and D. M. Filipovic, J. Phys. B **38**, 3489 (2005) and references therein.
[22] B. Predojevic, D. Sevic, V. Pejcev, B. P. Marinkovic and D. M. Filipovic, J. Phys. B **38**, 1329 (2005) and references therein.
[23] A. Z. Msezane, Z. Felfli, M. Ya. Amusia, Zhifan Chen and L. V. Chernysheva, Phys. Rev. A **65**, 054701 (2002)
[24] M. Ya. Amusia, L. V. Chernysheva, Z. Felfli, and A. Z. Msezane, i, Phys. Rev. A **65**, 62705 (2002)
[25] L. F. Zhu, H. D. Cheng, Z. S. Yuan, X. J. Liu, J. M. Sun, and K. Z. Xu, Phys. Rev. A **72** 042703 (2006),
[26] M. Ya. Amusia, L. V. Chernysheva, S. T. Manson, A. Z. Msezane and V. Radojević, Phys. Rev. Lett. **88**, 093002 (2002).
[27] L. V. Chernysheva, M. Ya. Amusia, Z. Felfli, and A. Z. Msezane, Abstracts of Contributed Papers, Vol. 2, XXIV ICPEAC (Rosario, Argentina, 2005) Eds. F.D. Colavecchia, P.D. Fainstein, J. Fiol, *et al*, page 252.
[28] M. Ya. Amusia, <u>Atomic Photoeffect</u> (Plenum Press, New York - London, 1990)
[29] Zhifan Chen and A.Z. Msezane, Phys. Rev. A **72** 050702(R) (2005); Zhifan Chen and A.Z. Msezane, J. Phys. B **39**, 4355 (2006)
[30] Zhifan Chen and A.Z. Msezane, Phys. Rev. A **68** 054701 (2003); Zhifan Chen and A.Z. Msezane, J. Phys. B **33**, 5397 (2000)

[31] M. Ya. Amusia, A. S. Baltenkov, L. V. Chernysheva, Z. Felfli, S. T. Manson and A. Z. Msezane, J. Phys. B **37**, 937 (2004) and references therein
[32] M. Ya. Amusia, N. A. Cherepkov, L. V. Chernysheva, Z. Felfli, and A. Z. Msezane, Phys. Rev. A **70**, 062709 (2004).
[33] A. Kivimäki, U. Hergenhahn, B. Kempgens, R. Hentges, M. N. Piancastelli, K. Maier, A. Ruedel, J. J. Tulkki, and A. M. Bradshaw, Phys. Rev. A **63**, 012716 (2001).
[34] O. Hemmers, R. Guillemmin, E. Kanter, B. Krässig, D. W. Lindle, S. H. Southworth, R. Wehlitz, J. Baker, A. Hudson, M. Lotrakul, D. Rolles, C. Stolte, C. Tran, A. Wolska, S. W. Yu, M. Ya. Amusia, K. T. Cheng, L. V. Chernysheva, W. R. Johnson and S. T. Manson, Phys. Rev. Lett. **91**, 053002 (2003)
[35] S. Ricz, R. Sankari, Á. Kövér, M. Jurvansuu, D. Varga, J. Nikkinen, T. Ricsoka, H. Aksela and S. Aksela, Phys. Rev. A **67**, 012712 (2003)
[36] M. Ya. Amusia, A. S. Baltenkov, L. V. Chernysheva, Z. Felfli, S. T. Manson and A. Z. Msezane, Phys. Rev. A **67**, 6070 (2003)
[37] L. D. Landau and E. M. Lifshitz, *Quantum Mechanics (Non-Relativistic Theory)*, Vol. 3 (Oxford: Butterworth-Heinemann, 1999)
[38] M. Ya. Amusia and L. V. Chernysheva, *Computation of Atomic Processes* (IOP Publishing Ltd, Bristol and Philadelphia, 1997)

Table 1

```
     GOS      for discrete dipole levels with  q=0.00001
   ---------------------------------------------------------
```

Ne

| E,RY | HF-L | HF-V | RPAE-L | RPAE-V | |
|---|---|---|---|---|---|
| 1.58893 | .2816E-01 | .2682E-01 | .2655E-01 | .3551E-01 | 2p-3d |
| 1.63788 | .1558E-01 | .1483E-01 | .1470E-01 | .1974E-01 | 2p-4d |
| 1.66057 | .8801E-02 | .8378E-02 | .8106E-02 | .1092E-01 | 2p-5d |

| E,RY | HF-L | HF-V | RPAE-L | RPAE-V | |
|---|---|---|---|---|---|
| 1.34815 | .1564E+00 | .1444E+00 | .1679E+00 | .1731E+00 | 2p-3s |
| 1.56349 | .2779E-01 | .2553E-01 | .2944E-01 | .3041E-01 | 2p-4s |
| 1.62788 | .9917E-02 | .9101E-02 | .1058E-01 | .1096E-01 | 2p-5s |

```
***********************************************************************
```

Ar

| E,RY | HF-L | HF-V | RPAE-L | RPAE-V | |
|---|---|---|---|---|---|
| .89702 | .2964E+00 | .2647E+00 | .3158E+00 | .3132E+00 | 3p-4s |
| 1.06277 | .5571E-01 | .4947E-01 | .3915E-01 | .3877E-01 | 3p-5s |
| 1.11637 | .2055E-01 | .1822E-01 | .1488E-01 | .1475E-01 | 3p-6s |
| 1.06778 | .1624E+00 | .9771E-01 | .1788E+00 | .1792E+00 | 3p-3d |
| 1.11821 | .8241E-01 | .4888E-01 | .8760E-01 | .8780E-01 | 3p-4d |
| 1.14136 | .4509E-01 | .2657E-01 | .4116E-01 | .4129E-01 | 3p-5d |

```
***********************************************************************
```

Kr                       without 4p-6s,4p-7s

| E,RY | HF-L | HF-V | RPAE-L | RPAE-V | |
|---|---|---|---|---|---|
| .93330 | .2673E+00 | .1525E+00 | .2614E+00 | .2607E+00 | 4p-4d |
| .98429 | .1316E+00 | .7385E-01 | .1263E+00 | .1258E+00 | 4p-5d |
| 1.00760 | .7116E-01 | .3964E-01 | .6358E-01 | .6329E-01 | 4p-6d |

without 4p-4d,4p-5d

| .78002 | .3752E+00 | .3364E+00 | .3576E+00 | .3454E+00 | 4p-5s |
| .93403 | .7049E-01 | .6282E-01 | .6592E-01 | .6463E-01 | 4p-6s |
| .98479 | .2621E-01 | .2332E-01 | .2351E-01 | .2313E-01 | 4p-7s |

```
***********************************************************************
```

Xe

| E,RY | HF-L | HF-V | RPAE-L | RPAE-V | |
|---|---|---|---|---|---|
| .67262 | .4185E+00 | .3589E+00 | .4263E+00 | .4043E+00 | 5p-6s |
| .80827 | .8236E-01 | .7005E-01 | .1252E+00 | .1190E+00 | 5p-7s |
| .85442 | .3130E-01 | .2656E-01 | .4608E-01 | .4376E-01 | 5p-8s |

| E,RY | HF-L | HF-V | RPAE-L | RPAE-V | |
|---|---|---|---|---|---|
| .79777 | .4839E+00 | .2483E+00 | .4573E+00 | .4444E+00 | 5p-5d |
| .84984 | .2309E+00 | .1160E+00 | .2123E+00 | .2057E+00 | 5p-6d |
| .87351 | .1235E+00 | .6145E-01 | .1069E+00 | .1032E+00 | 5p-7d |

```
********************************************************
```

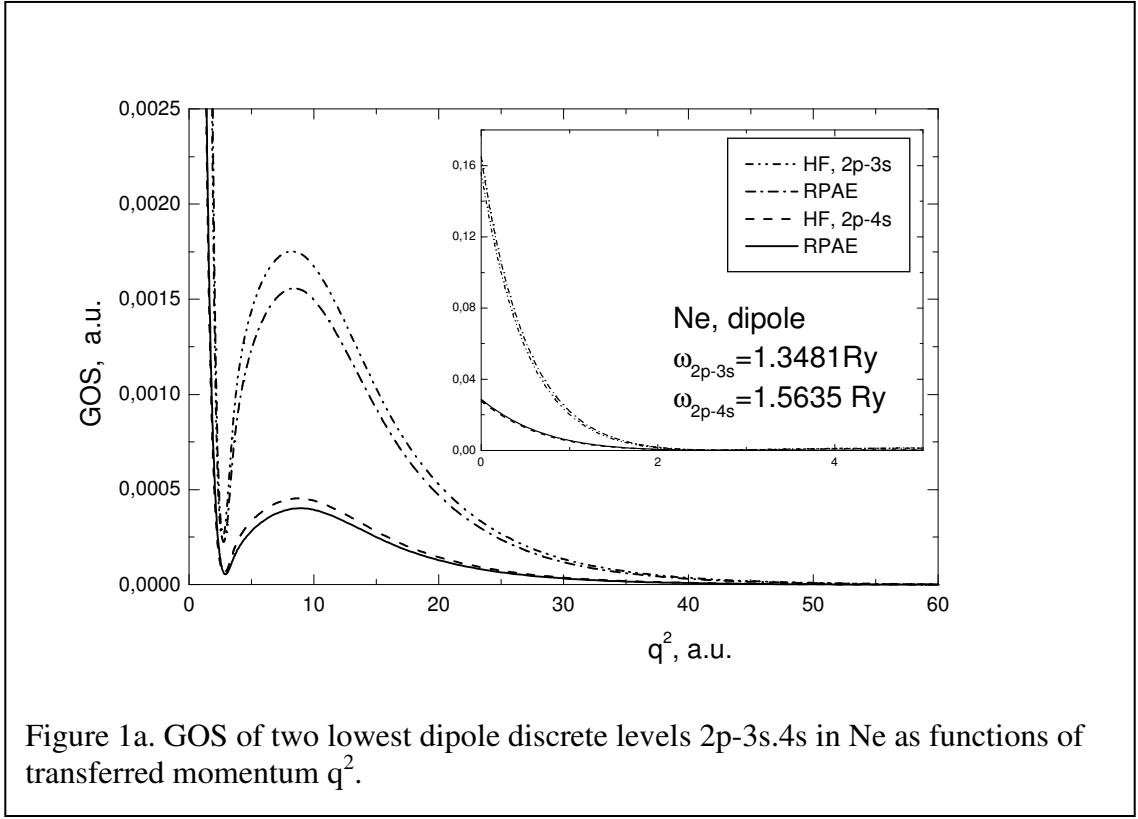

Figure 1a. GOS of two lowest dipole discrete levels 2p-3s.4s in Ne as functions of transferred momentum $q^2$.

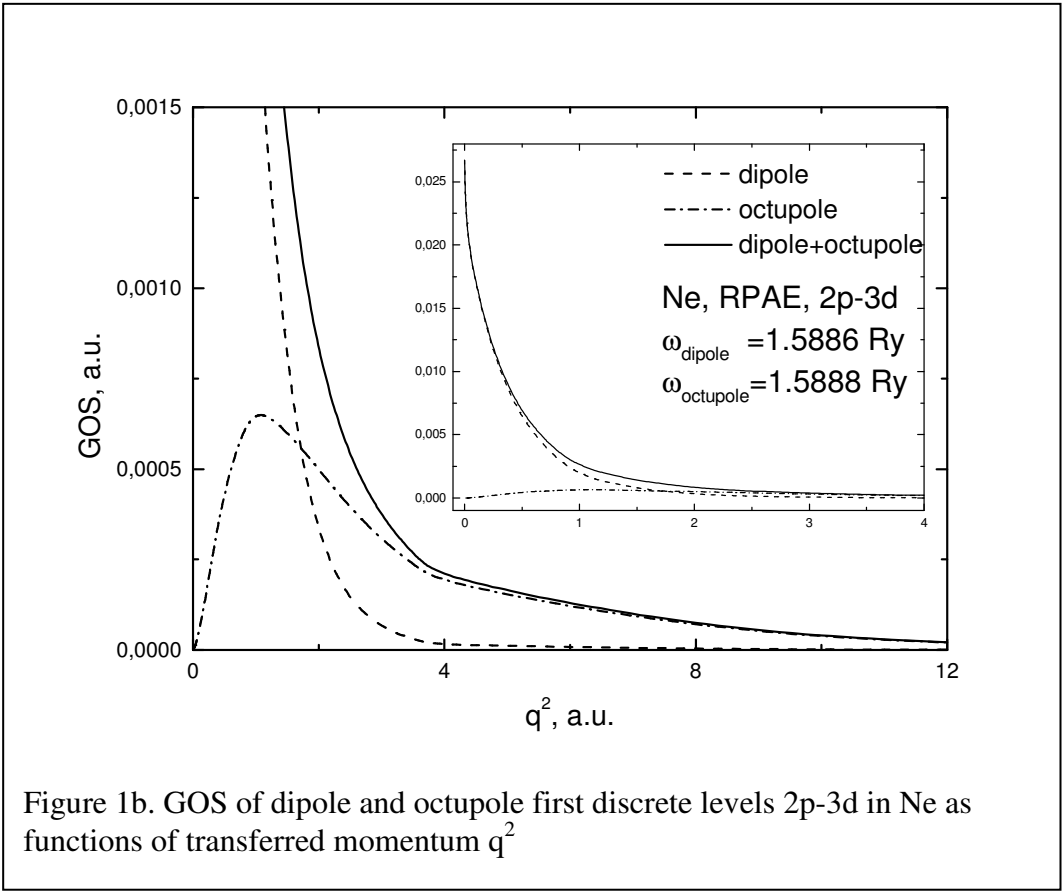

Figure 1b. GOS of dipole and octupole first discrete levels 2p-3d in Ne as functions of transferred momentum $q^2$

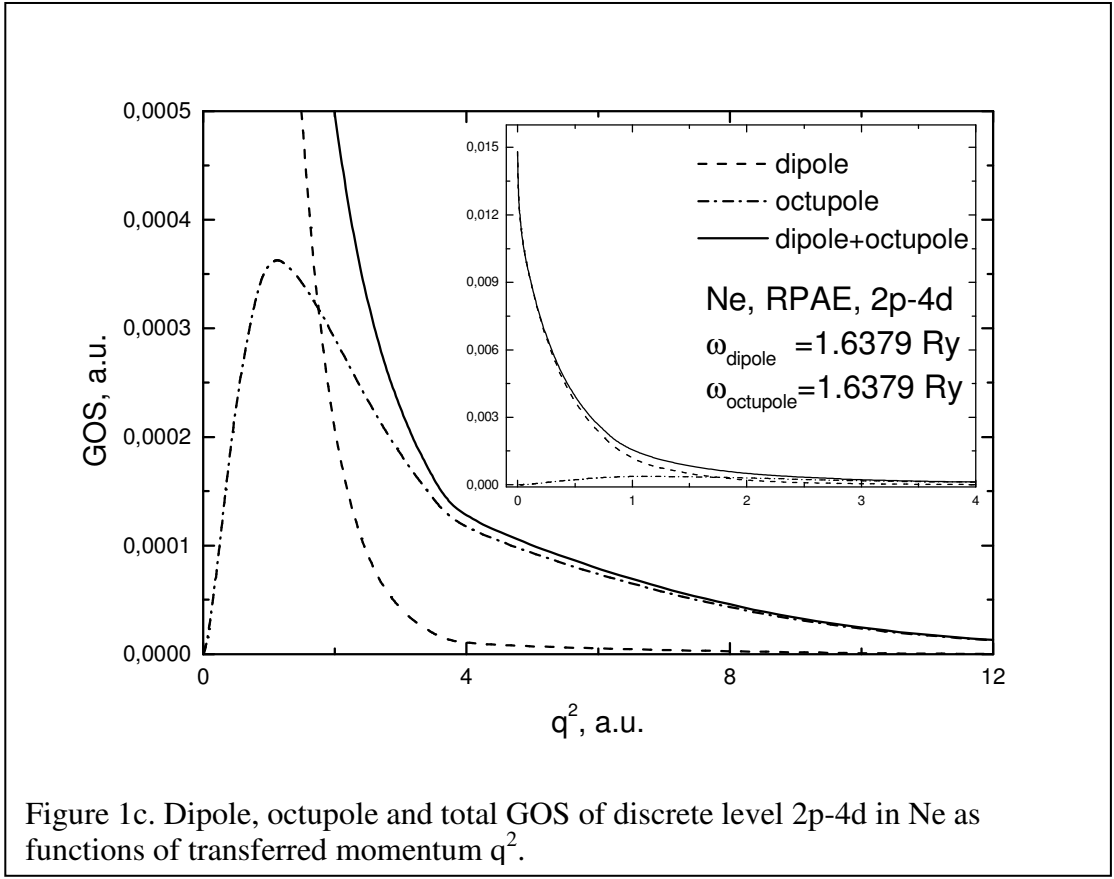

Figure 1c. Dipole, octupole and total GOS of discrete level 2p-4d in Ne as functions of transferred momentum $q^2$.

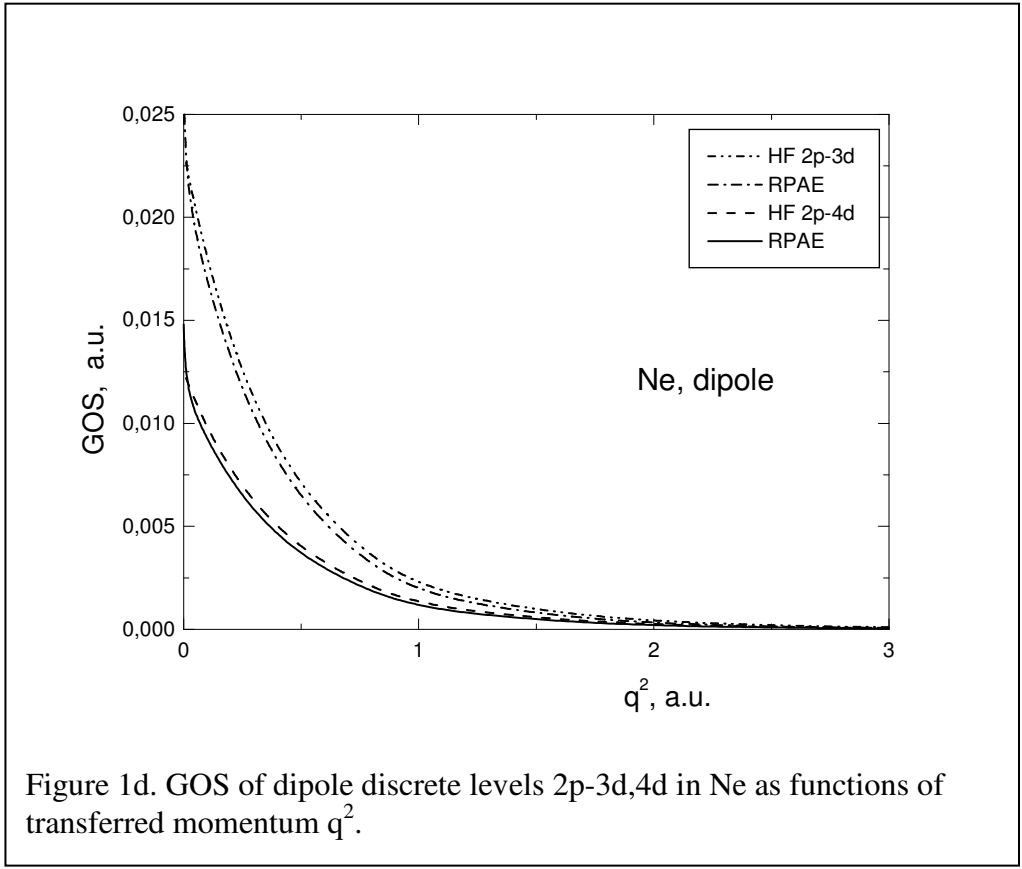

Figure 1d. GOS of dipole discrete levels 2p-3d,4d in Ne as functions of transferred momentum $q^2$.

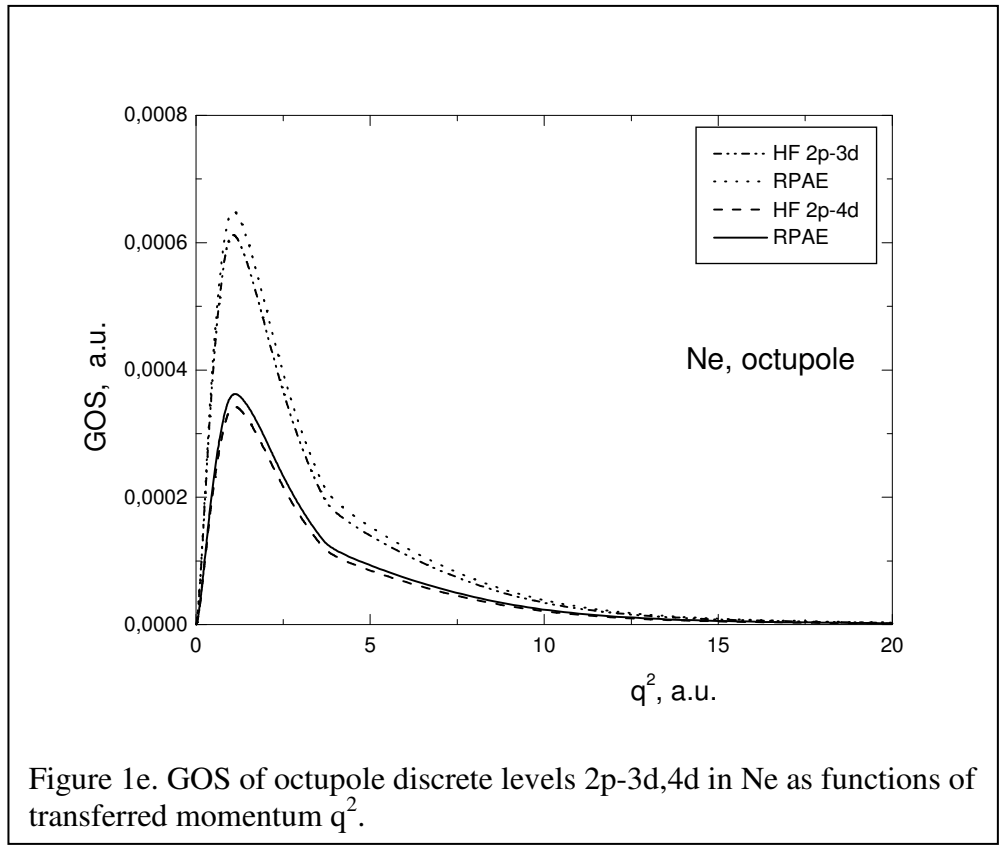

Figure 1e. GOS of octupole discrete levels 2p-3d,4d in Ne as functions of transferred momentum $q^2$.

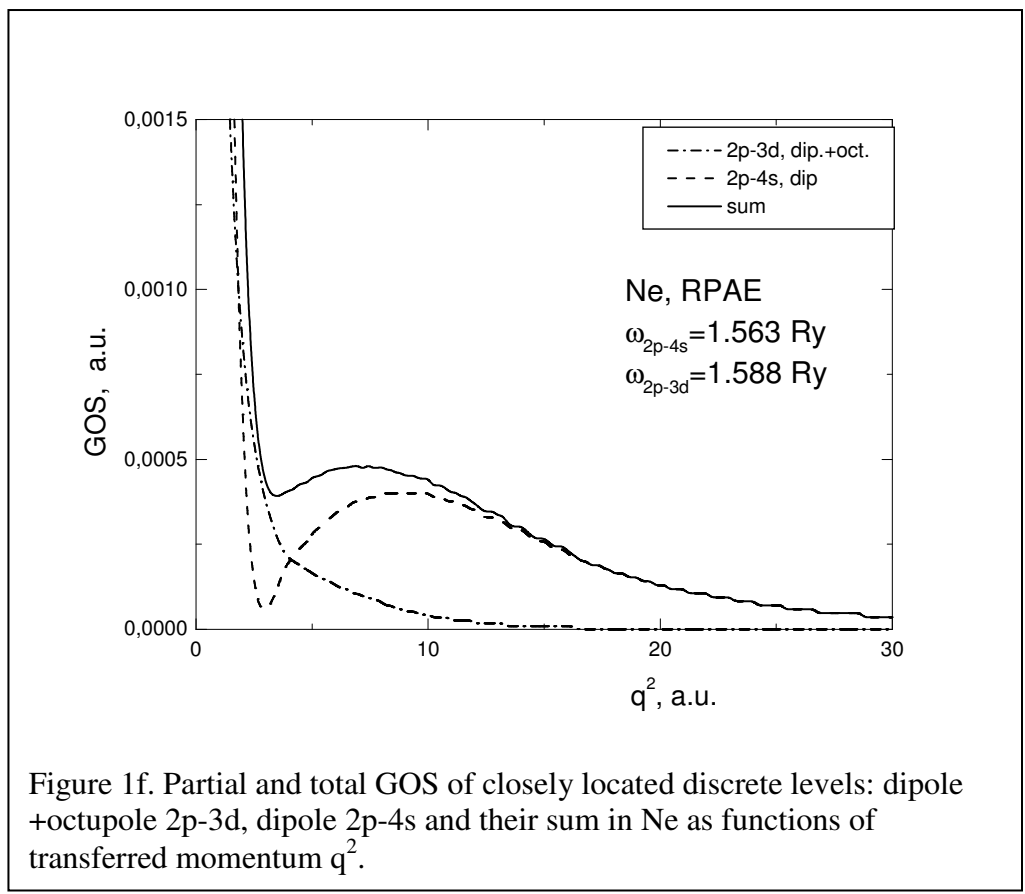

Figure 1f. Partial and total GOS of closely located discrete levels: dipole +octupole 2p-3d, dipole 2p-4s and their sum in Ne as functions of transferred momentum $q^2$.

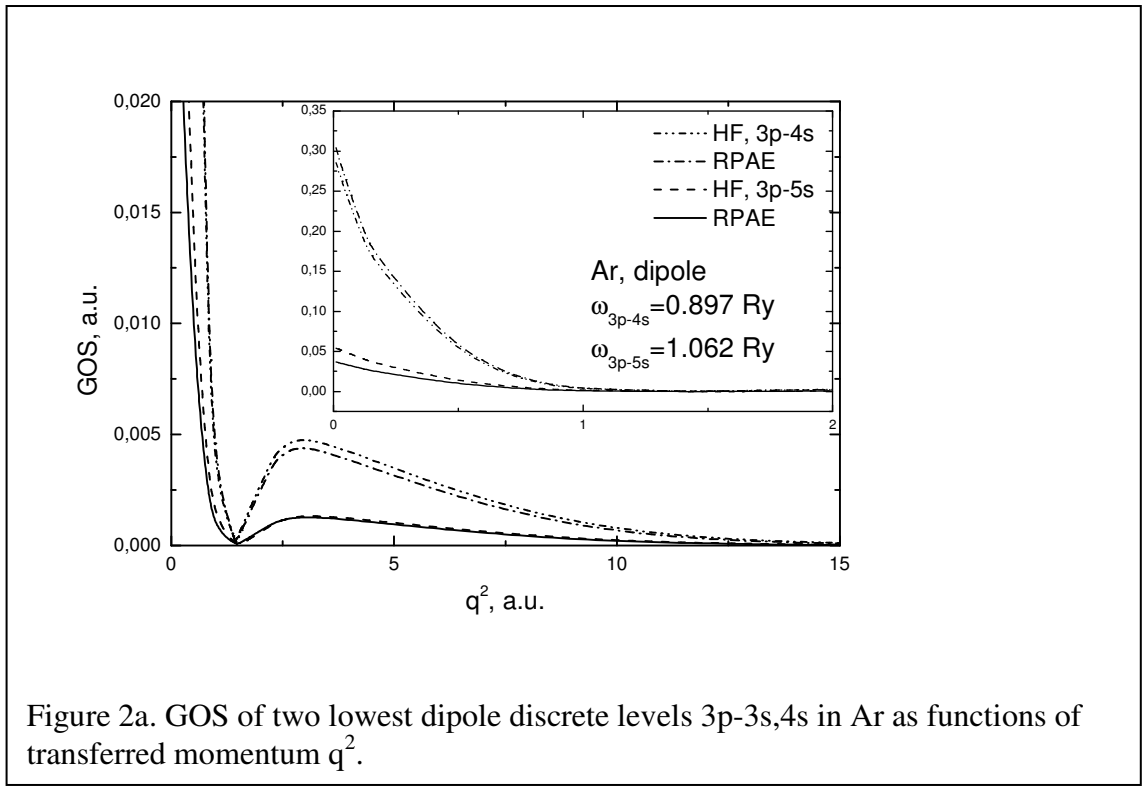

Figure 2a. GOS of two lowest dipole discrete levels 3p-3s,4s in Ar as functions of transferred momentum $q^2$.

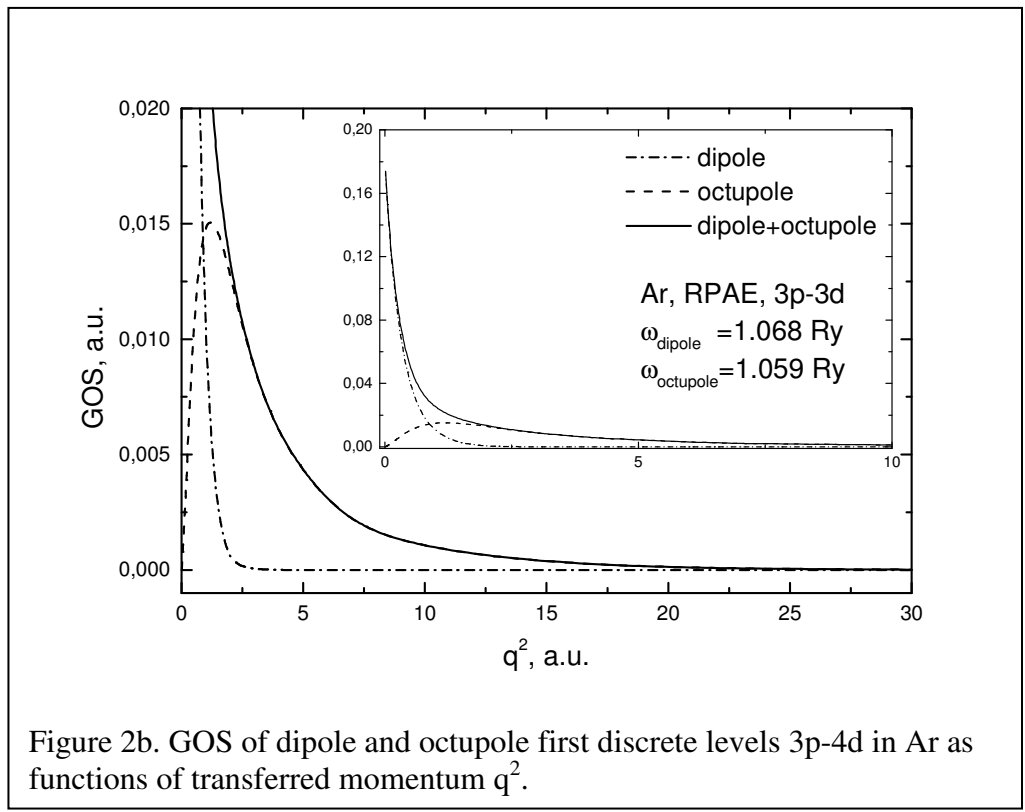

Figure 2b. GOS of dipole and octupole first discrete levels 3p-4d in Ar as functions of transferred momentum $q^2$.

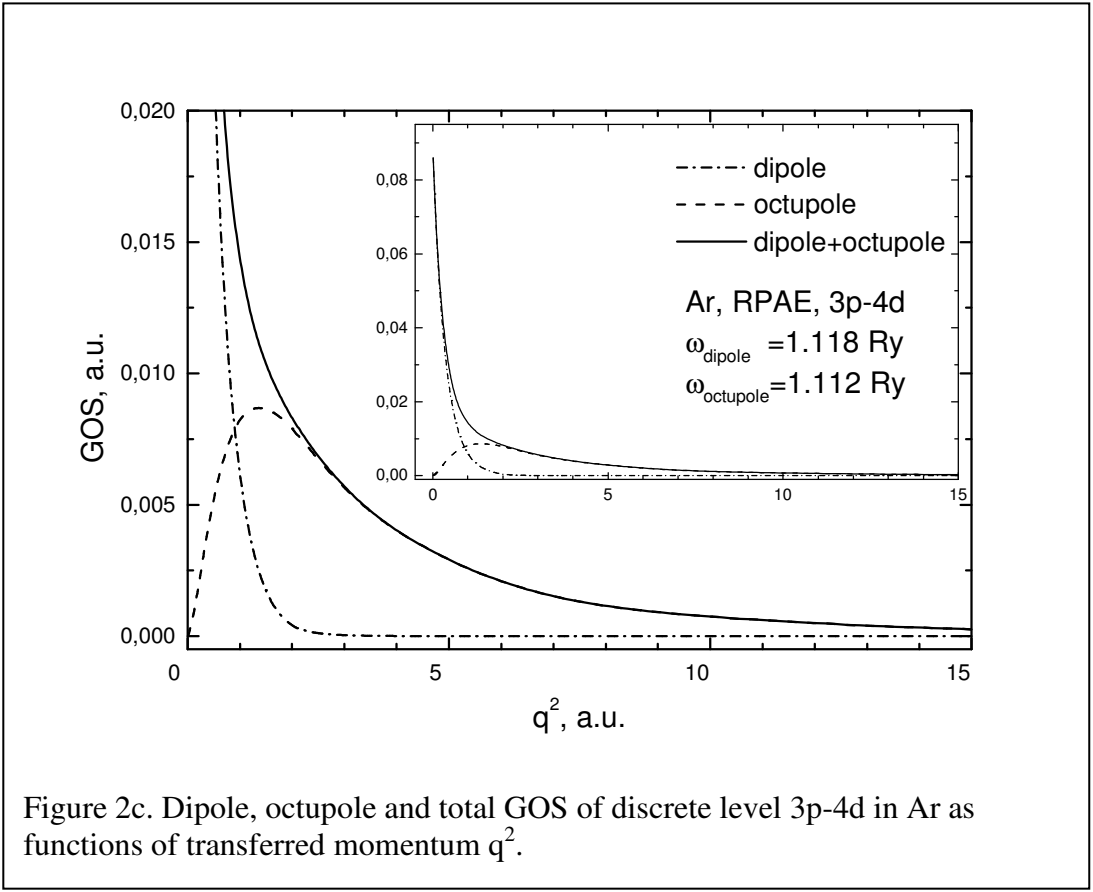

Figure 2c. Dipole, octupole and total GOS of discrete level 3p-4d in Ar as functions of transferred momentum $q^2$.

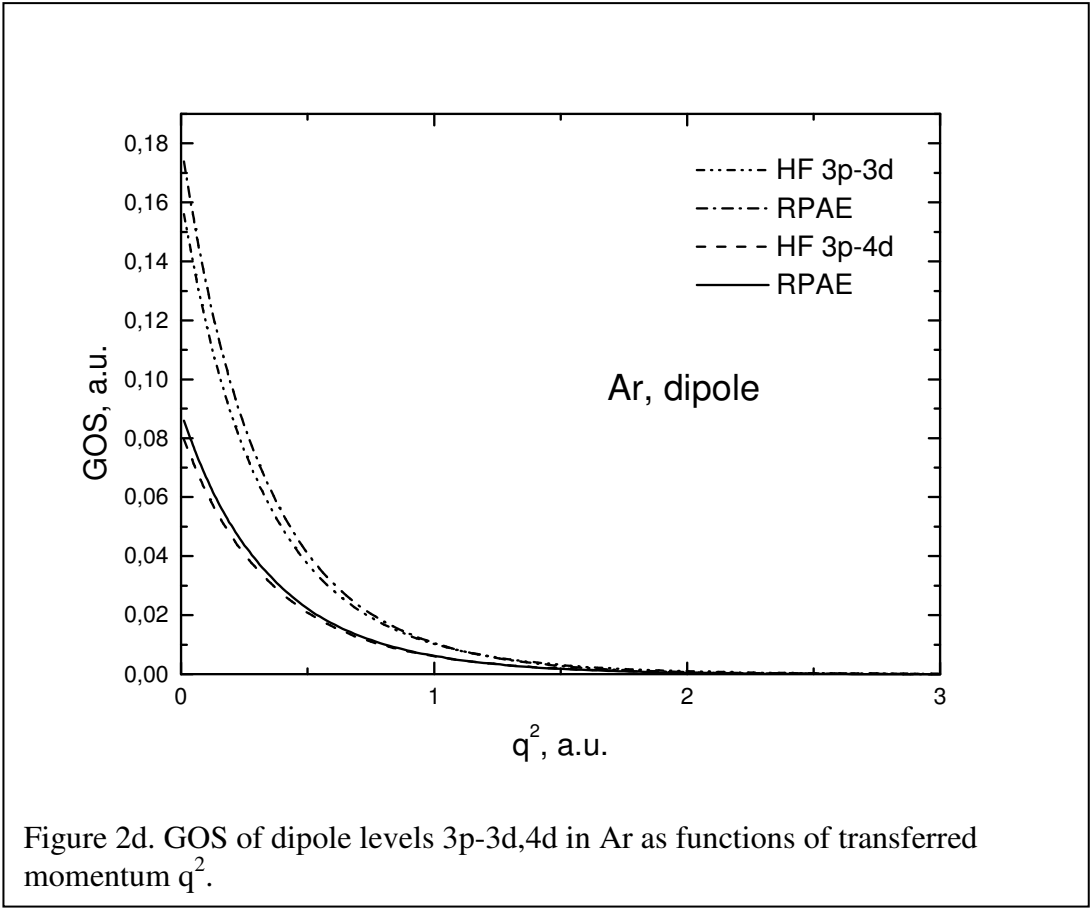

Figure 2d. GOS of dipole levels 3p-3d,4d in Ar as functions of transferred momentum $q^2$.

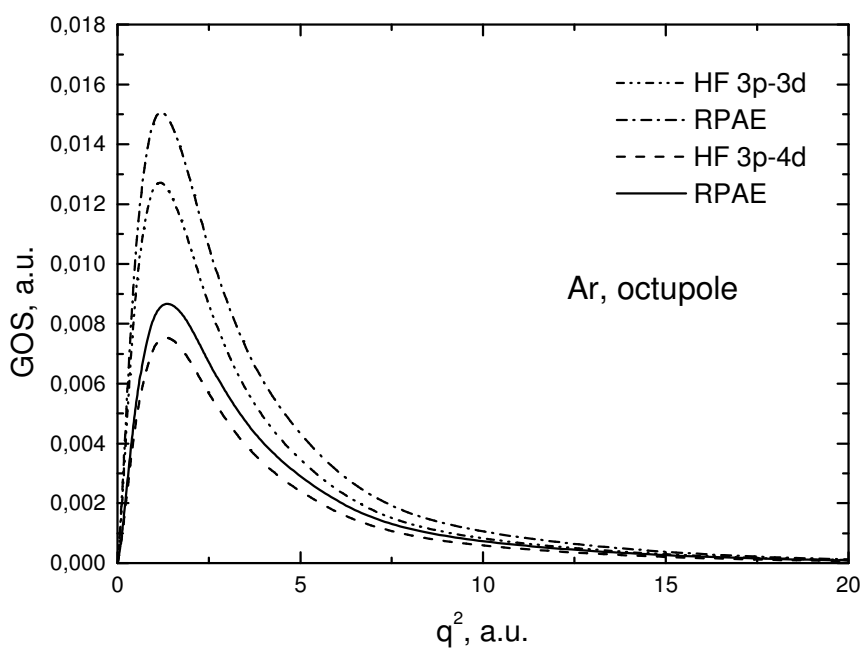

Figure 2e. . GOS of octupole levels 3p-3d,4d in Ar as functions of transferred momentum $q^2$.

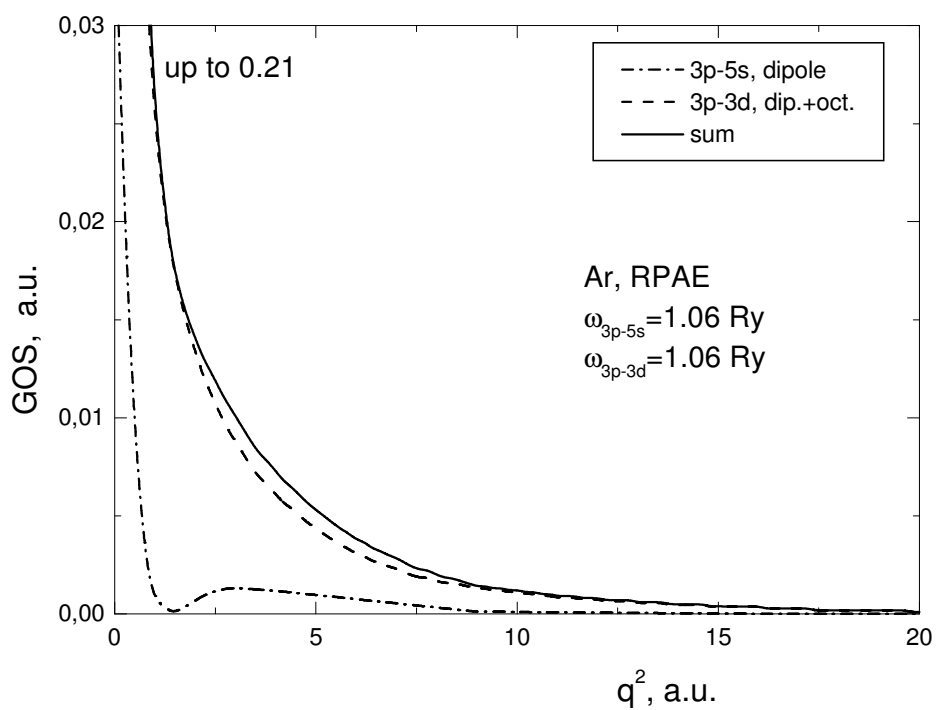

Figure 2f. Partial and total GOS of closely located discrete levels: dipole +octupole 3p-3d, dipole 3p-5s and their sum in Ar as functions of transferred momentum $q^2$.

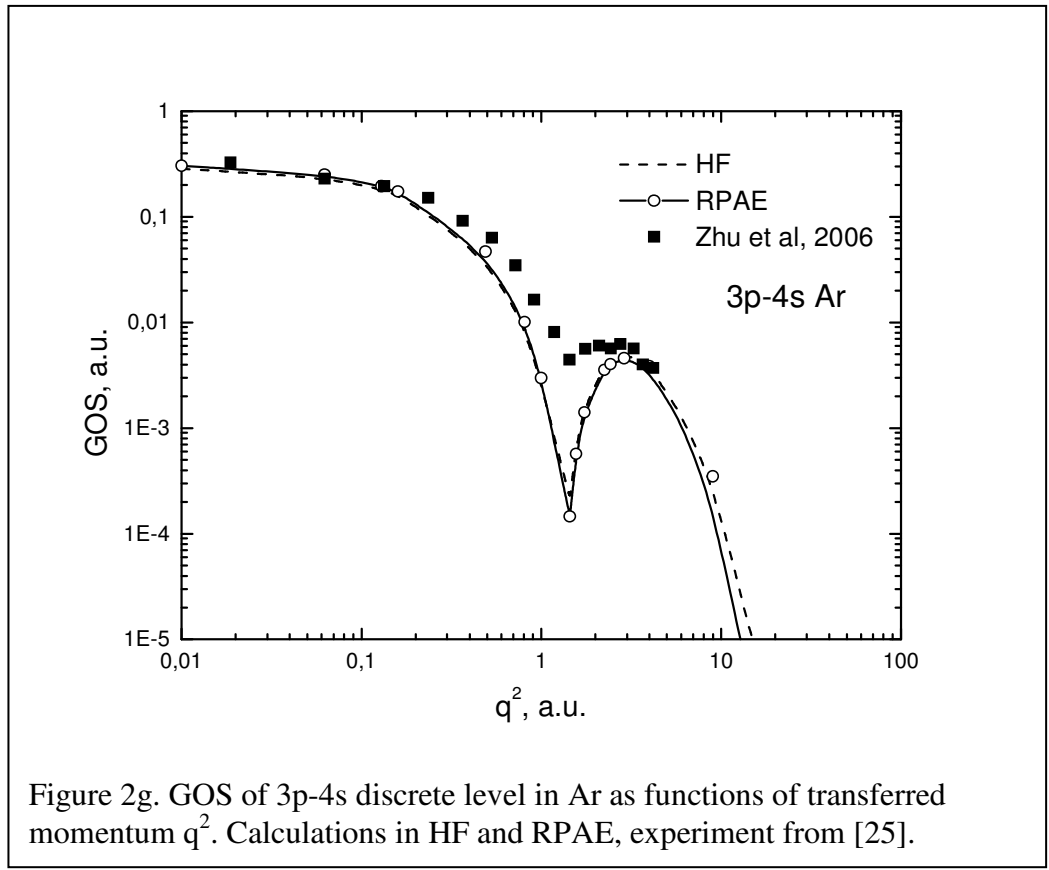

Figure 2g. GOS of 3p-4s discrete level in Ar as functions of transferred momentum $q^2$. Calculations in HF and RPAE, experiment from [25].

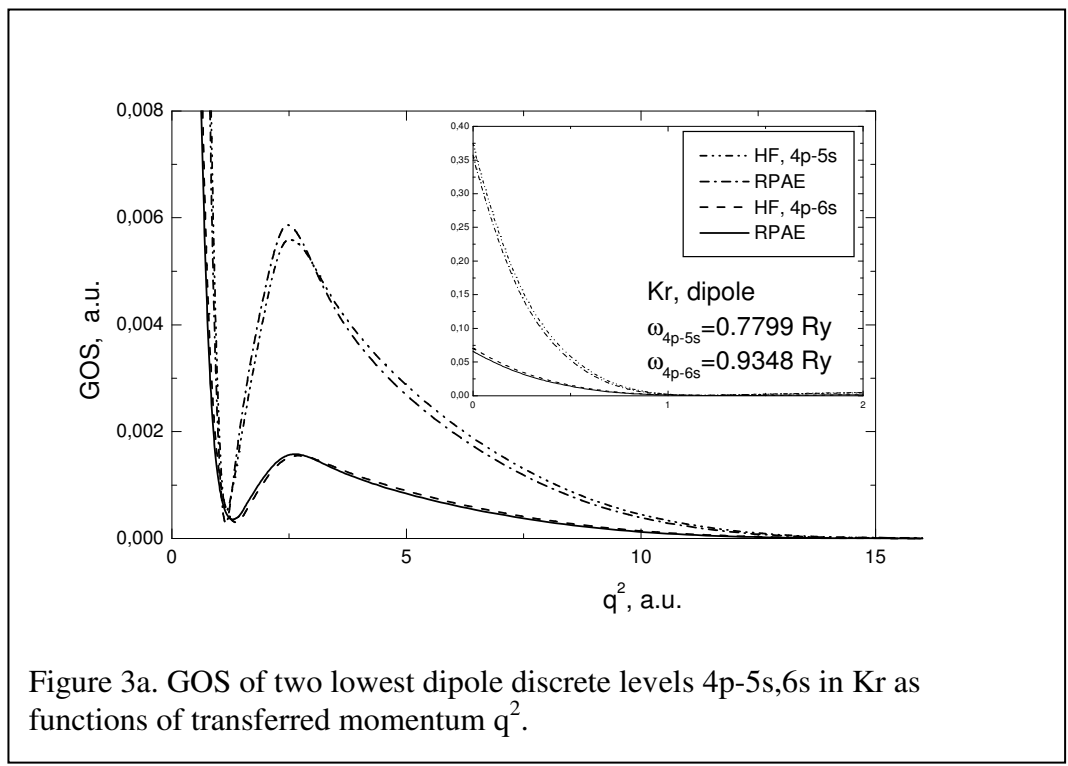

Figure 3a. GOS of two lowest dipole discrete levels 4p-5s,6s in Kr as functions of transferred momentum $q^2$.

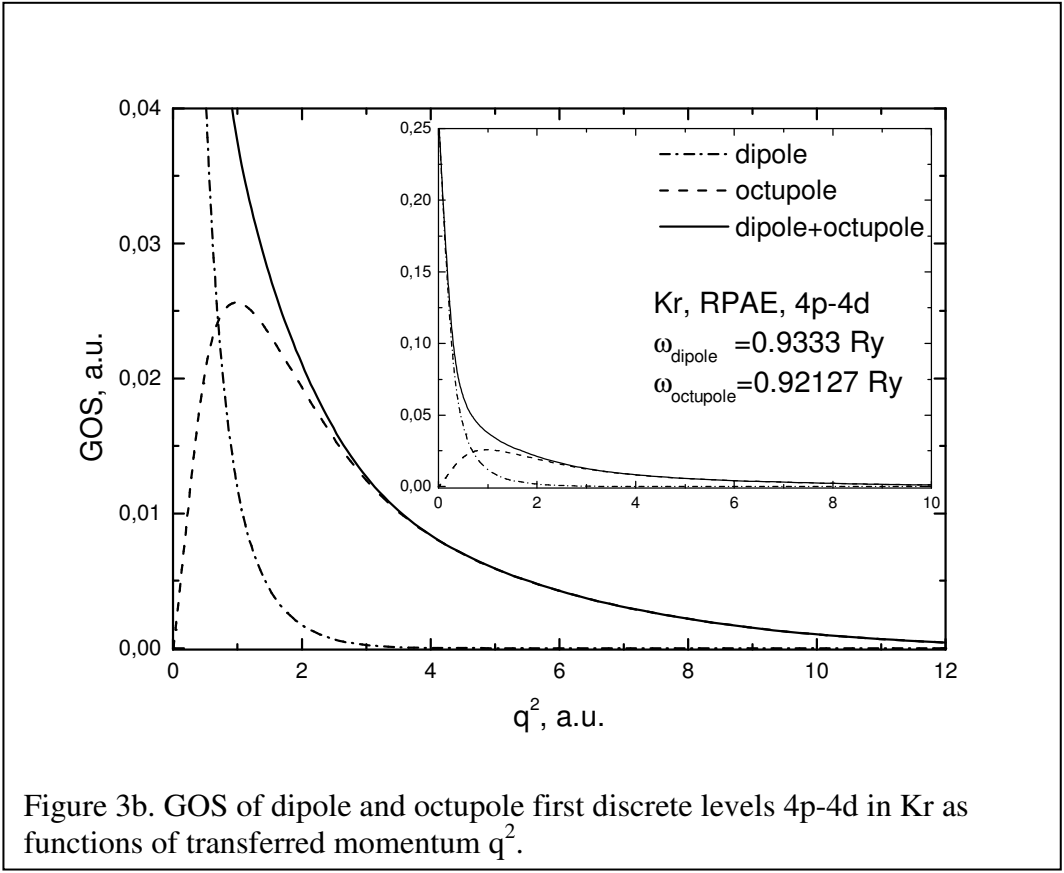

Figure 3b. GOS of dipole and octupole first discrete levels 4p-4d in Kr as functions of transferred momentum $q^2$.

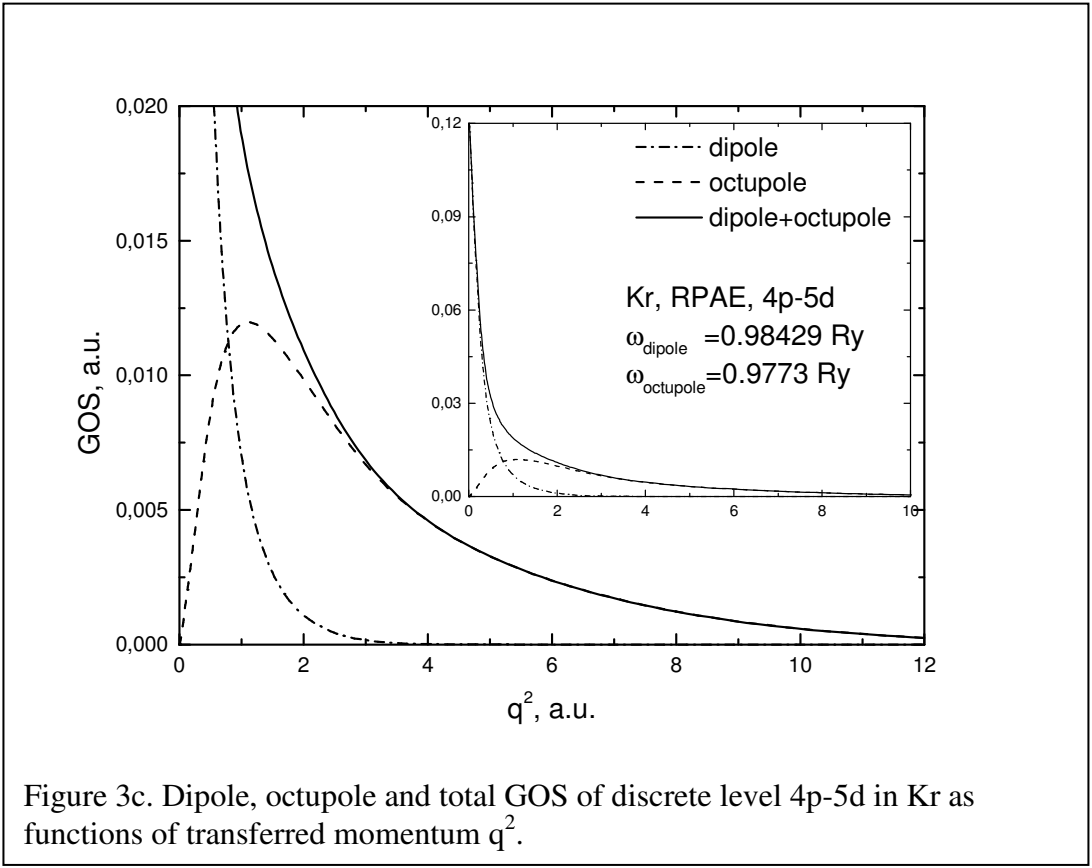

Figure 3c. Dipole, octupole and total GOS of discrete level 4p-5d in Kr as functions of transferred momentum $q^2$.

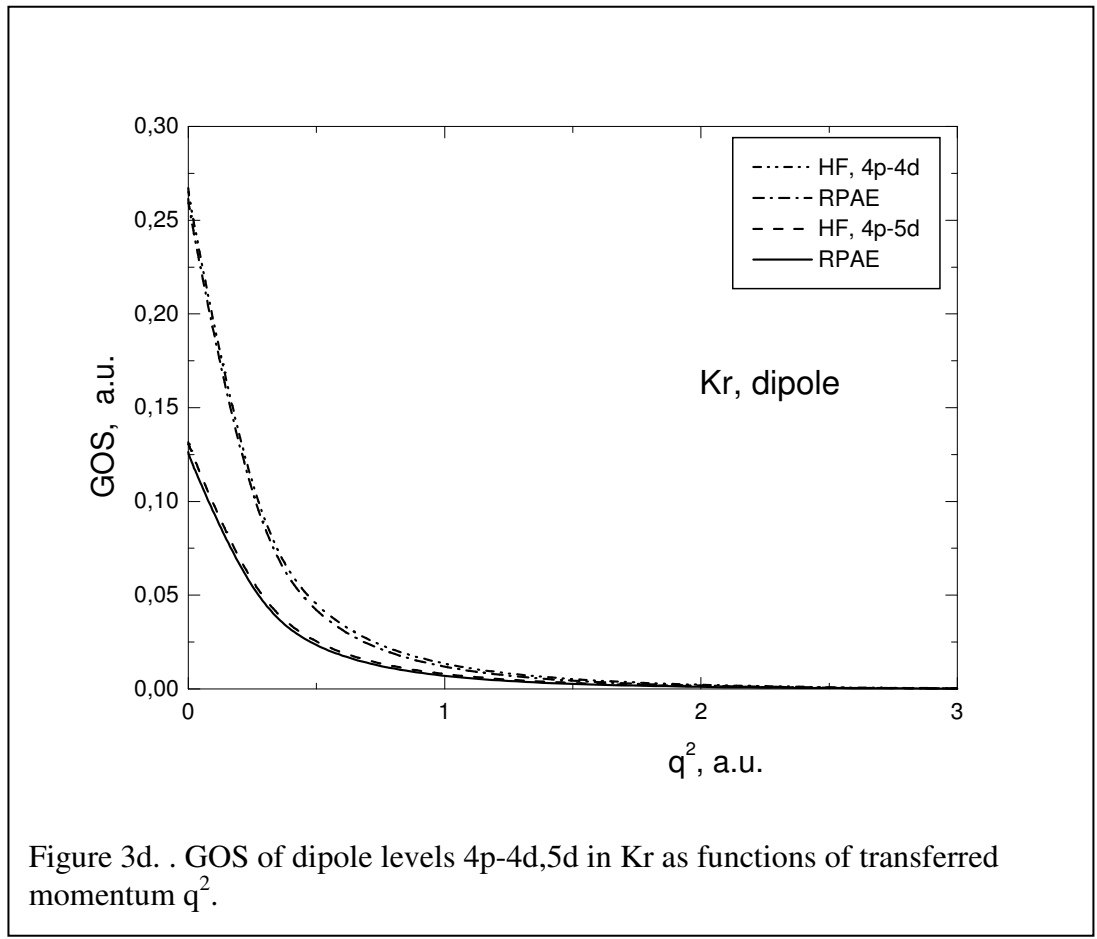

Figure 3d. . GOS of dipole levels 4p-4d,5d in Kr as functions of transferred momentum $q^2$.

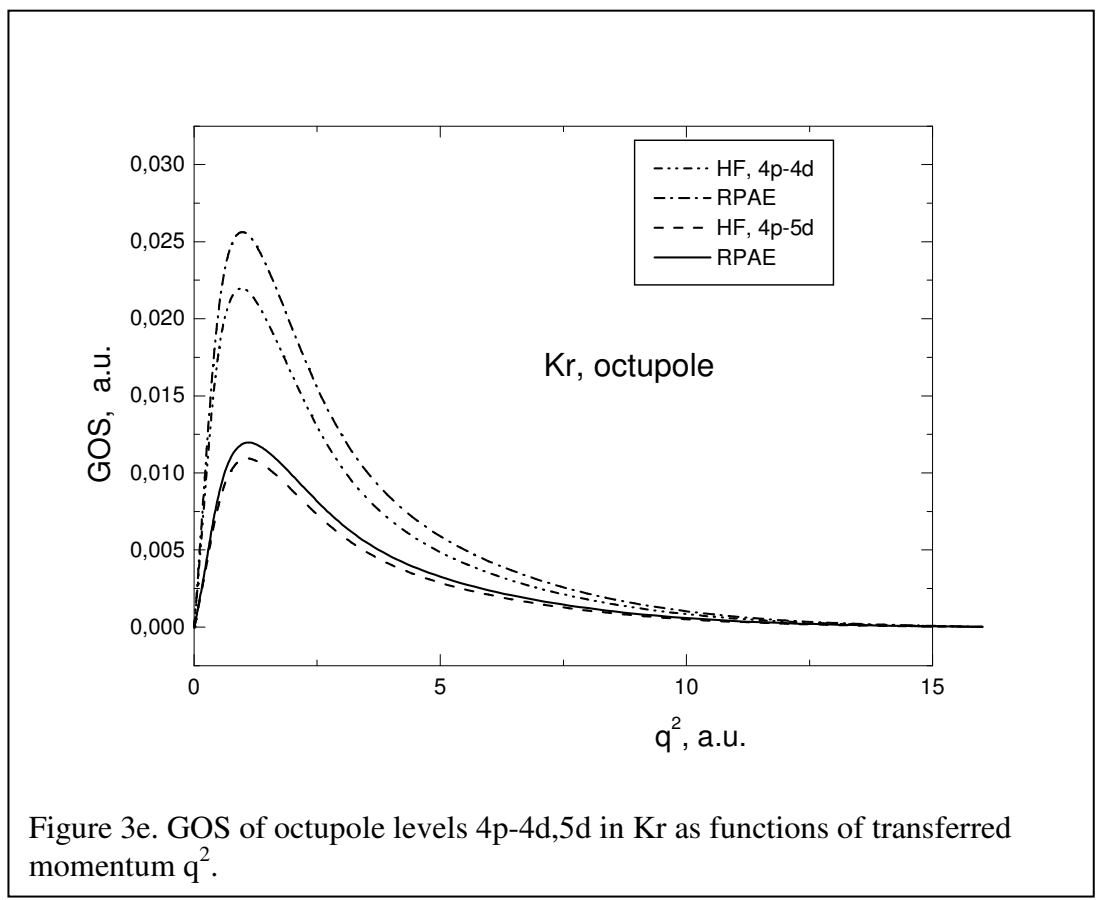

Figure 3e. GOS of octupole levels 4p-4d,5d in Kr as functions of transferred momentum $q^2$.

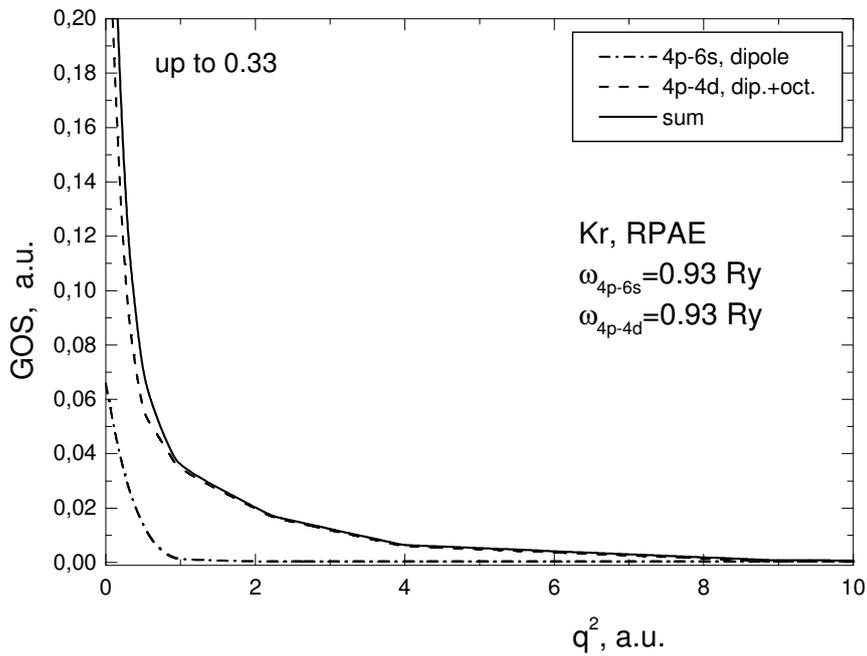

Figure 3f. Partial and total GOS of closely located discrete levels: dipole +octupole 4p-4d, dipole 4p-6s and their sum in Kr as functions of transferred momentum $q^2$.

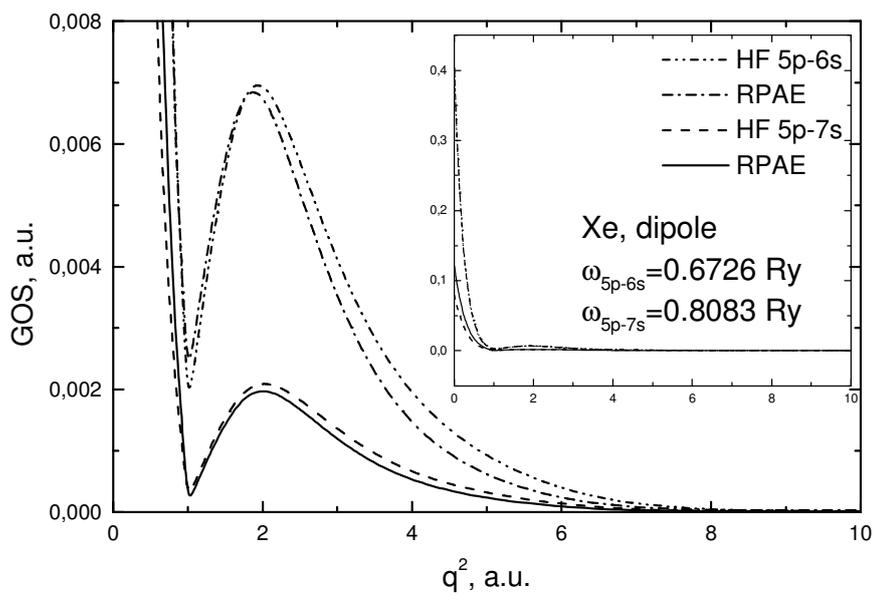

Figure 4a. GOS of two lowest dipole discrete levels 5p-6s,7s in Xe as functions of transferred momentum $q^2$.

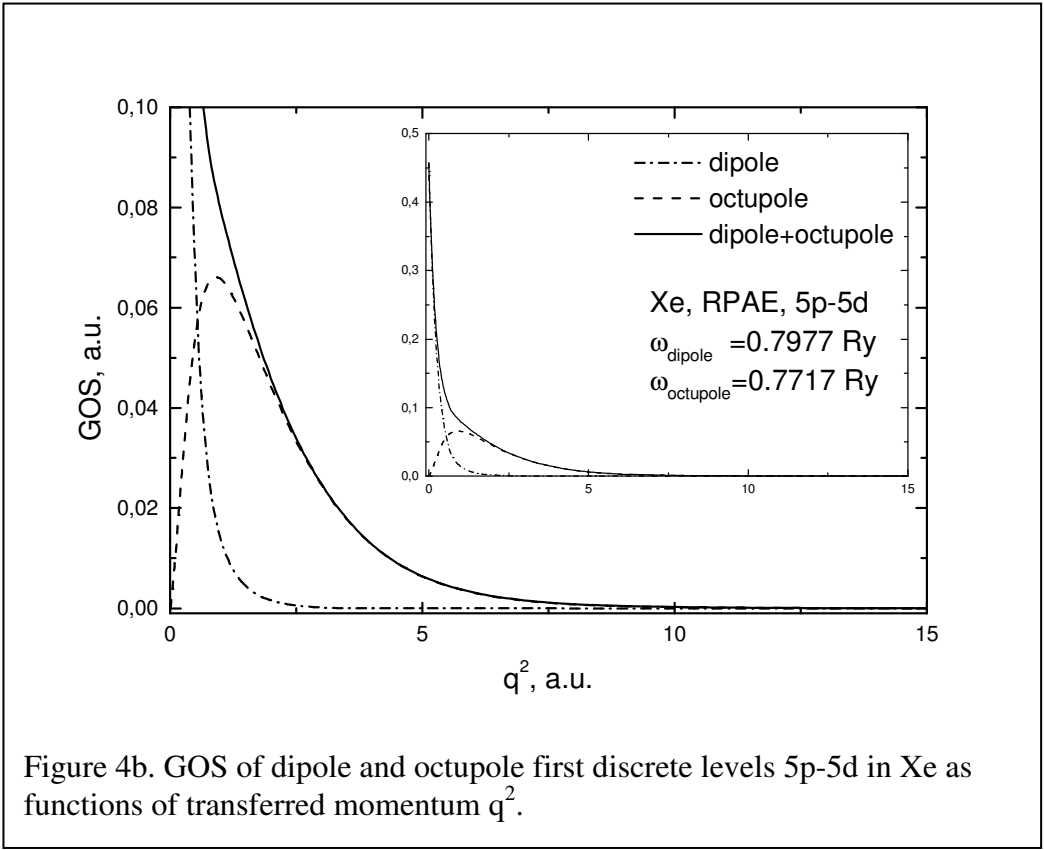

Figure 4b. GOS of dipole and octupole first discrete levels 5p-5d in Xe as functions of transferred momentum $q^2$.

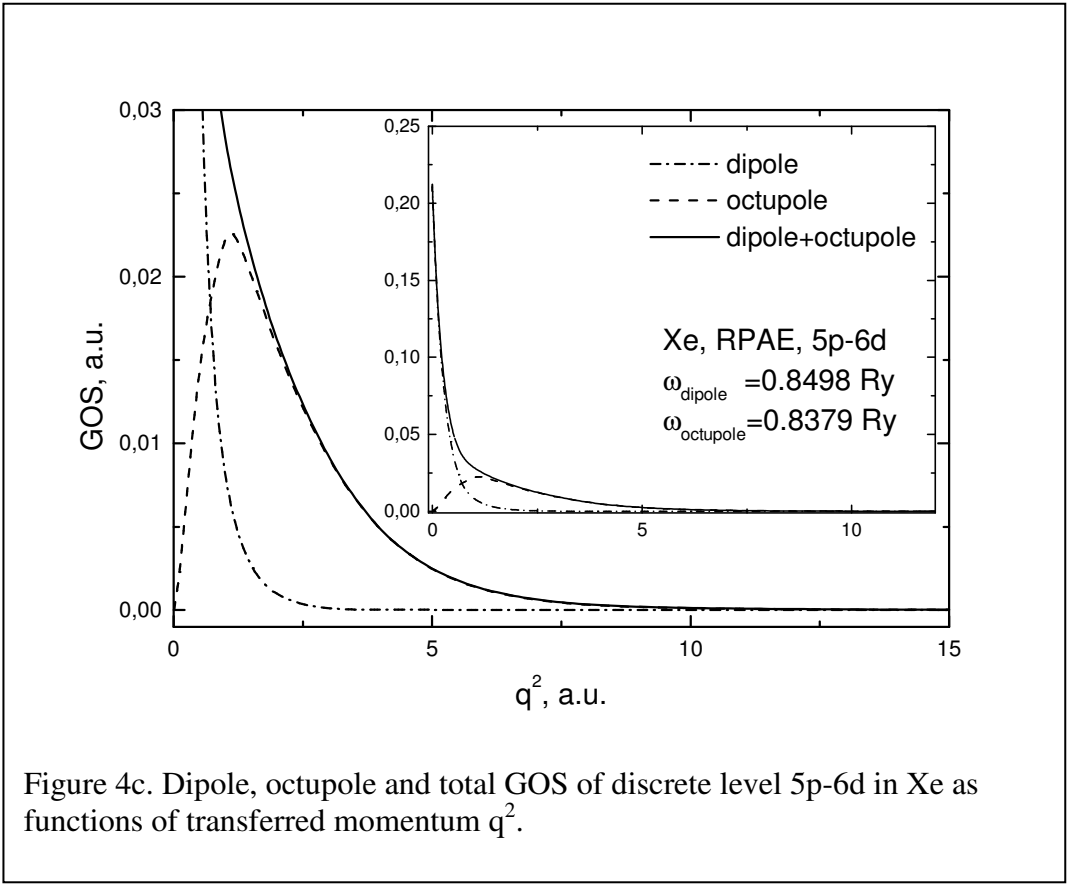

Figure 4c. Dipole, octupole and total GOS of discrete level 5p-6d in Xe as functions of transferred momentum $q^2$.

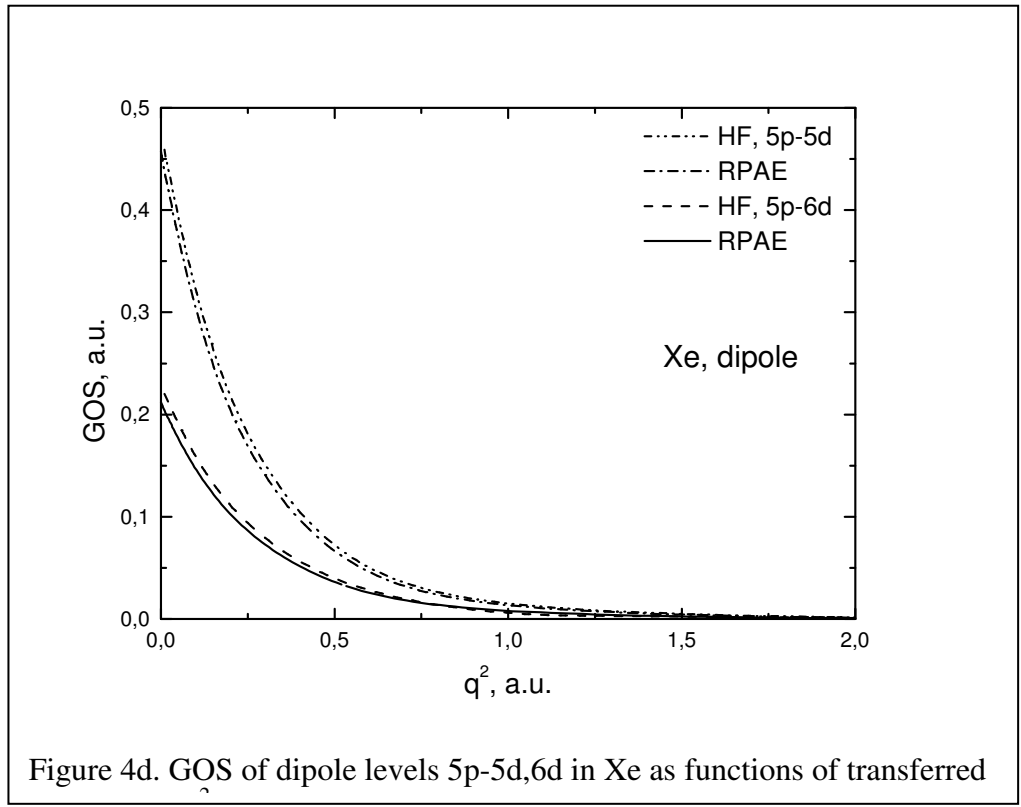

Figure 4d. GOS of dipole levels 5p-5d,6d in Xe as functions of transferred

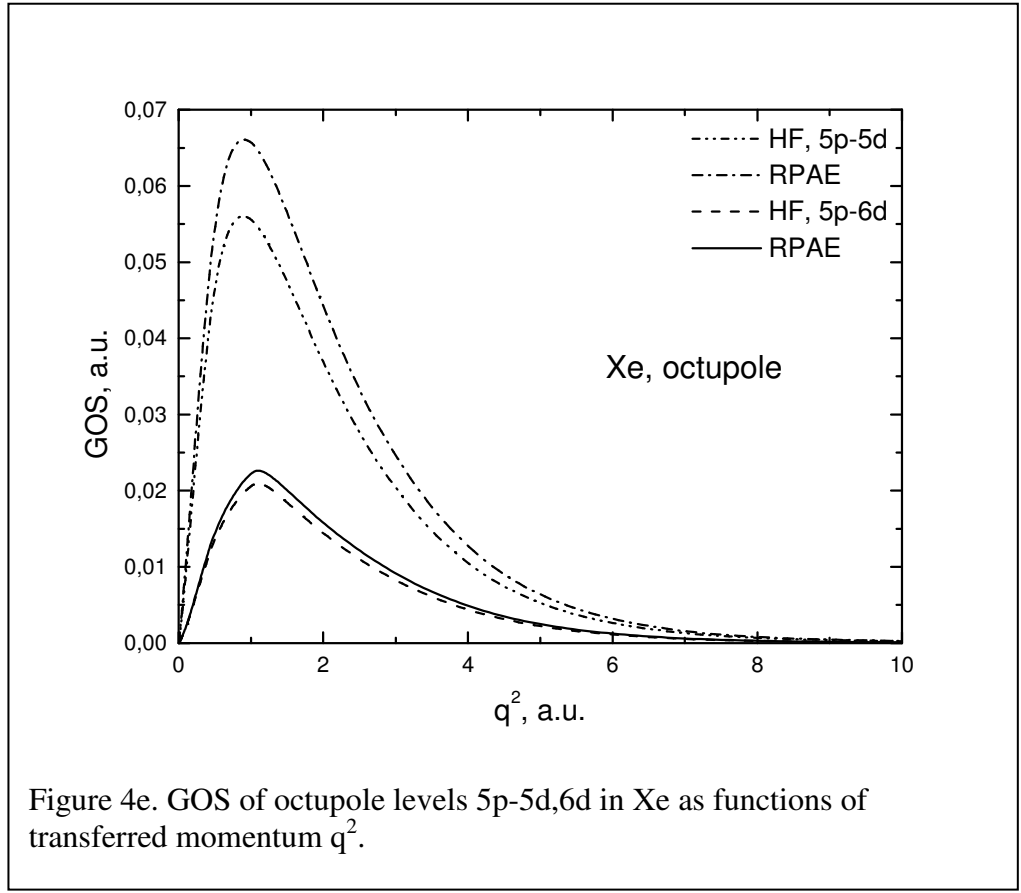

Figure 4e. GOS of octupole levels 5p-5d,6d in Xe as functions of transferred momentum $q^2$.

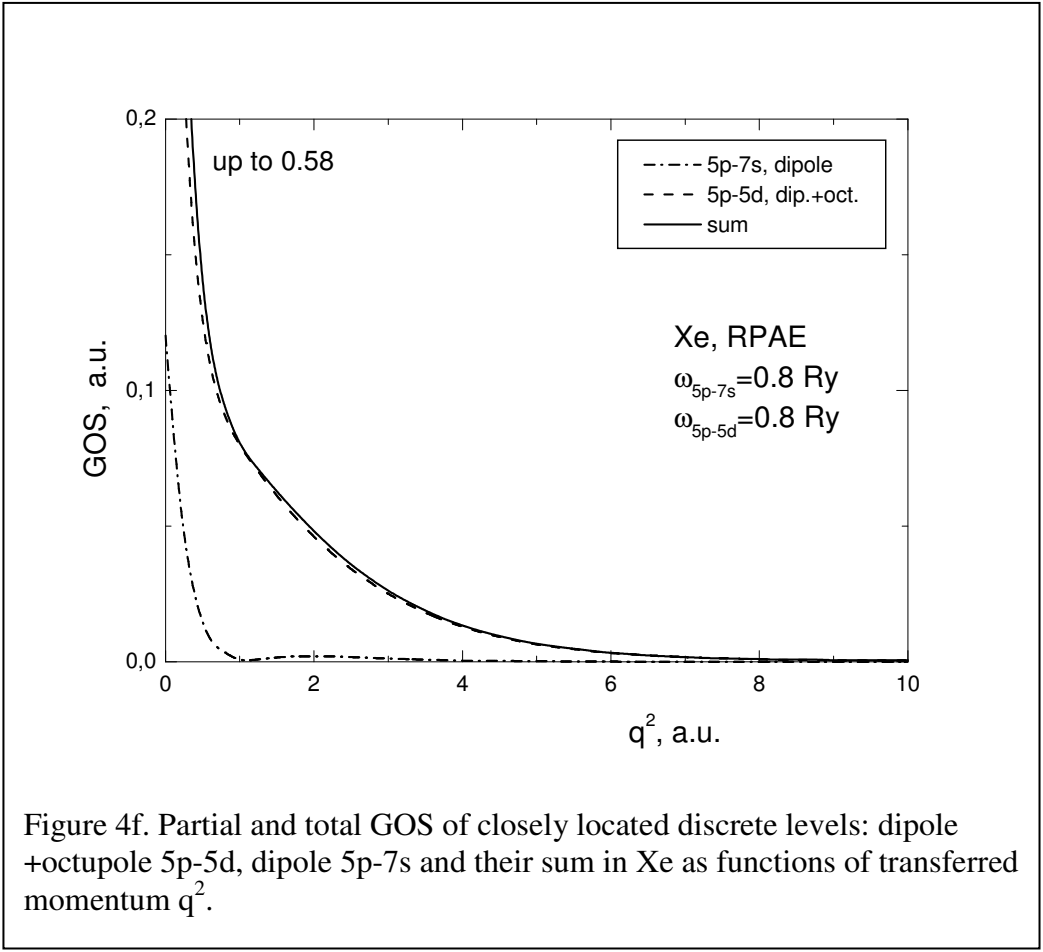

Figure 4f. Partial and total GOS of closely located discrete levels: dipole +octupole 5p-5d, dipole 5p-7s and their sum in Xe as functions of transferred momentum $q^2$.